\documentclass{nature}
\usepackage{epsfig,color,amssymb}
\usepackage{subfigure}
\usepackage{amsfonts}
\usepackage{amscd}
\usepackage{amsmath}
\usepackage{multirow}
\usepackage{chemarrow}
\usepackage{dcolumn}
\usepackage{bm}
\usepackage{graphicx}
\usepackage{enumerate}
\usepackage{epsfig}
\usepackage{subfigure}
\usepackage{xcolor}
\usepackage{multirow}
\usepackage{ulem}
\usepackage{braket}
\usepackage{comment}
\usepackage{enumitem}
\usepackage{amsthm}
\renewcommand{\emph}[1]{\textit{#1}}

\begin{document}

\title{Measurement-Device-Independent Twin-Field Quantum Key Distribution}

\author{Hua-Lei Yin$^{1,2}$ \& Yao Fu$^{2}$}

\maketitle

\begin{affiliations}
\item
National Laboratory of Solid State Microstructures and School of Physics, Nanjing University, Nanjing 210093, China
\item
Department of Physics, Zhejiang Institute of Modern Physics and ZJU-Phoenix Synergetic Innovation Center in Quantum Technology, Zhejiang University, Hangzhou 310027, China\\
Correspondence and requests for materials should be addressed to H.-L.Y. (email: hlyin@nju.edu.cn)
\end{affiliations}

\baselineskip24pt



\begin{abstract}
The ultimate aim of quantum key distribution (QKD) is improving the transmission distance and key generation speed. Unfortunately, it is believed to be limited by the secret-key capacity of quantum channel without quantum repeater.
Recently, a novel twin-field QKD (TF-QKD) is proposed to break through the limit, where the key rate is proportional to the square-root of channel transmittance.
Here, by using the vacuum and one-photon state as a qubit, we show that the TF-QKD can be regarded as a measurement-device-independent QKD (MDI-QKD) with single-photon Bell state measurement. Therefore, the MDI property of TF-QKD can be understood clearly. Importantly, the universal security proof theories can be directly used for TF-QKD, such as BB84 encoding, six-state encoding and reference-frame-independent scheme. Furthermore, we propose a feasible experimental scheme for the proof-of-principle experimental demonstration.
\end{abstract}

\maketitle
Throughout history, the battle between encryption and decryption never ends. Currently, relying on computational complexity, the widely used public-key cryptosystem becomes vulnerable to quantum computing attacks.
The one-time pad is the only provably secure cryptosystem according to information theory known today. Thereinto, an important issue exists that the
common secret key is at least as long as the message itself and can be used only once.
Quantum key distribution (QKD) constitutes the only way to solve the real time key distribution problem\cite{bennett1984quantum}. QKD allows two distant parties to establish a string of secret keys with information-theoretic security\cite{Scarani:2009:The,Weedbrook:2012:Gaussian}. One can ensure legitimate parties to exchange messages with perfect confidentiality by combining QKD with one-time pad.

The longest transmission distance of QKD has been implemented over 421 km with ultralow-loss optical fiber\cite{boaron2018secure} and 1200 km satellite-to-ground\cite{liao2017satellite}. Improving the transmission distance and key rate are the most important tasks of QKD research. However, this task has been proven impossible beyond a certain limit without quantum repeaters\cite{takeoka2014fundamental,pirandola2017fundamental}. The secret-key capacity of quantum channel can be used to bound the extractable maximum secret key\cite{takeoka2014fundamental,pirandola2017fundamental}. Generally, the secret-key capacity can be regarded as a linear key rate Pirandola-Laurenza-Ottaviani-Banchi (PLOB) bound\cite{pirandola2017fundamental} $R_{\textrm{PLOB}}=-\log_{2}(1-\eta)$, where $\eta$ is the transmittance.
To overcome the rate-distance limit of QKD, quantum repeaters are usually believed as a strong candidate\cite{duan2001long,Sangouard:2011:Quantum}. However, the long-time quantum memory and high-fidelity entanglement distillation are far from feasible. Despite the recent advance\cite{azuma2015all} relaxing the requirement, the actual implementation is also difficult to realize, for example, quantum non-demolition (QND) measurement. Although the trusted relay-based QKD has been deployed over 2000 km\cite{qiu2014quantum}, its security is compromised.

Recently, a novel protocol called twin-field QKD (TF-QKD)\cite{lucamarini2018overcoming} has been proposed to overcome the rate-distance limit. The secret key rate of TF-QKD has been scaled with the square-root of the transmittance, $R\sim O(\sqrt{\eta})$. In the TF-QKD, a pair of optical fields are generated respectively at locations of two remote parties and then sent to the untrusted center to implement single-photon detection. Compared with measurement-device-independent QKD (MDI-QKD)\cite{lo2012measurement}, TF-QKD retains the properties of being immune to all detector attack, multiplexing of expensive single-photon detectors and natural star network architecture.
In the original paper of TF-QKD\cite{lucamarini2018overcoming}, the communication parties, Alice and Bob, prepare the phase-randomized coherent state with phase encoding in $X$ and $Y$ basis. To acquire the correction of raw keys, they should announce the random phase of each pulse. The key rate of unconditional security proof is still missing in the original paper\cite{lucamarini2018overcoming}. Various different important works have been shown to give the key rate formulas with information-theoretic security\cite{tamaki2018information,ma2018phase,Wang2018Sending,cui:2018:phase,curty:2018:simple,Lin:2018:A}.

Here, we prove that TF-QKD can be seen as a special type of MDI-QKD. Thereinto, a qubit is physically implemented by a two-dimensional subspace with vacuum and one-photon state.
One can consider that the untrusted center performs the single-photon Bell state measurement (BSM) while Alice and Bob prepare quantum state in the complementary bases. Since the vacuum state is immune to the loss, it can always have a detection (detector without click means a successful detection), thus the probability of coincident detection is exactly equal to that of single detection. Therefore, the TF-QKD inherits all positive features of MDI-QKD and increases the key rate a lot to break through the linear key rate bound. The unconditional security proof technologies with entanglement purification\cite{lo1999unconditional,shor2000simple}, information theory analysis\cite{kraus2005lower}, entropy uncertainty relation\cite{tomamichel2011uncertainty} can be directly applied in the TF-QKD.
The bit of $Z$ basis is independent of the phase misalignment. Naturally, there is no need to publish random phase of $Z$ basis and the state can be seen as a mixture of photon number states. Therefore, the distilled secret key of $Z$ basis in the TF-QKD can exploit the tagging-method of Gottesman-Lo-L\"{u}tkenhaus-Preskill (GLLP) analysis\cite{gottesman2004Security}.
Combining the decoy-state method\cite{hwang2003quantum,wang2005beating,lo2005decoy}, we could acquire the tight key rate formula of TF-QKD with BB84 encoding\cite{bennett1984quantum}, six-state encoding\cite{Lo2001Proof} and reference-frame-independent (RFI)\cite{laing2010reference} scheme .

\section*{Results}
\begin{figure*}
\centering
\includegraphics[width=14cm]{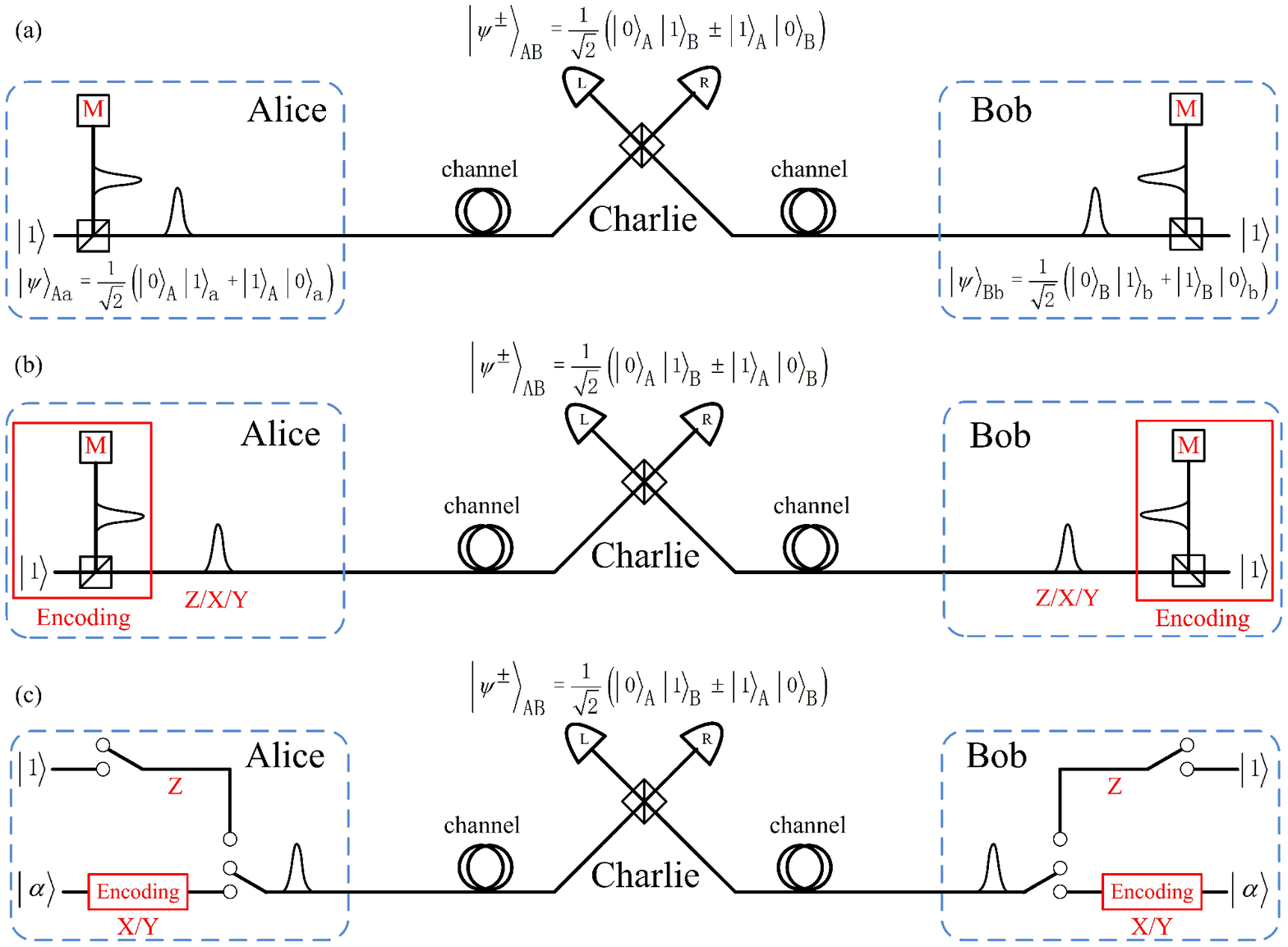}
\caption{Scheme to overcome the PLOB bound of QKD. (a) Setup for entanglement-based MDI-QKD with single-photon BSM.
Alice and Bob prepare single-photon Bell state, while Charlie implements entanglement swapping. $M$ represents the measurement operation, such as $Z$, $X$ and $Y$ basis. Alice and Bob implement the $M$ measurement operation after Charlie performs the single-photon BSM.
(b) Prepare-and-measure MDI-QKD with single-photon BSM. Alice and Bob directly prepare the qubit with superpositions of the vacuum and one-photon states.
Alice and Bob implement the $M$ measurement operation before Charlie performs the single-photon BSM.
(c) Effective TF-QKD with single-photon and laser sources. The photons from single-photon source and laser source are indistinguishable in every degree of freedom. The phase-reference of long-distance should be stabilized to implement laser interference. The single-photon source is used to implement $Z$ basis encoding, while the laser source is used to implement the phase encoding, such as $X$ and $Y$ basis.
} \label{f1}
\end{figure*}

\textbf{MDI-QKD with single-photon BSM.}
Here, let us first introduce an entanglement-based MDI-QKD with single-photon BSM protocol, as shown in Fig. \ref{f1}(a). Let $\{\ket{0}, \ket{1}\}$ represent $Z$ basis, where $\ket{0}$ and $\ket{1}$ are the vacuum and the one-photon state, respectively. Accordingly, the eigenvectors of $X$ basis and $Y$ basis are $\ket{\pm}=(\ket{0}\pm\ket{1})/\sqrt{2}$ and $\ket{\pm i}=(\ket{0}\pm i\ket{1})/\sqrt{2}$. Considering that one photon inputs a lossless symmetric beam splitter, the output state is a single-photon entangled state, $\ket{\psi^{+}}=(\ket{0}\ket{1}+\ket{1}\ket{0})/\sqrt{2}$.
Alice and Bob prepare a series of entangled states $\ket{\psi^{+}}_{Aa}=(\ket{0}_{A}\ket{1}_{a}+\ket{1}_{A}\ket{0}_{a})/\sqrt{2}$ and $\ket{\psi^{+}}_{Bb}=(\ket{0}_{B}\ket{1}_{b}+\ket{1}_{B}\ket{0}_{b})/\sqrt{2}$, respectively,
where $A$ ($B$) and $a$ ($b$) are a pair of field modes.
Afterwards, they hold the qubit of $a$ and $b$ modes and send the quantum states of $A$ and $B$ modes to the untrusted third party, Charlie, who performs the BSM to identify the two single-photon Bell states $\ket{\psi^{+}}_{AB}=(\ket{0}_{A}\ket{1}_{B}+\ket{1}_{A}\ket{0}_{B})/\sqrt{2}$ and $\ket{\psi^{-}}_{AB}=(\ket{0}_{A}\ket{1}_{B}-\ket{1}_{A}\ket{0}_{B})/\sqrt{2}$.
Therefore, a coincidence detection with $L$ click and $R$ no click  indicates a projection into the Bell
state $\ket{\psi^{+}}_{AB}$. A coincidence detection with $R$ click and $L$ no click, implies a projection into the Bell
state $\ket{\psi^{-}}_{AB}$. Note that the identification of any one Bell state is enough to prove the security.
When Charlie performs a successful BSM, the qubit that the legitimate users hold becomes a single-photon Bell state, the process of which can be regarded
as an entanglement swapping, as experimentally demonstrated\cite{sciarrino2002delayed}. Alice and Bob can utilize
quantum memory to store their qubit $a$ and $b$ modes. After Charlie announces the events through public channels whether
he has obtained a Bell state and which Bell state he has identified, Alice and Bob will measure qubit $a$ and $b$ modes, respectively.
They publish the basis information through an authenticated classical channel.
Bob will apply a bit flip when they choose $Z$ ($X$ or $Y$) basis and Charlie receives a Bell state $\ket{\psi^{\pm}}_{AB}$ $\left(\ket{\psi^{-}}_{AB}\right)$.
They use the data of $Z$ basis to form the raw key, while the data of other bases are all used to estimate the leaked information. Alice and Bob can acquire the secure key through the error correction and privacy amplification.

We can equivalently convert our entanglement-based protocol in Fig. \ref{f1}(a) to the prepare-and-measure protocol as shown in Fig. \ref{f1}(b) by the Shor-Preskill's arguments\cite{shor2000simple}. Let Alice and Bob measure the modes $a$ and $b$ before they send the qubit of $A$ and $B$ modes to Charlie, meaning Alice and Bob directly prepare the quantum state $A$ mode and $B$ mode. Other steps are all same to the entanglement-based protocol, including the BSM, basis comparison, bit flip, error correction and privacy amplification.
Hereafter, we use the TF state to represent the joint quantum state of Alice's $A$ mode and Bob's $B$ mode.
In the case of ideal detector (photon-number-resolving and without dark count) and lossless channel, the MDI-QKD with single-photon BSM protocol is similar with the two-photon BSM protocol. However, the single-photon BSM exploits the vacuum state identification, namely, detector without click, the case of TF state with $\ket{1}_{A}\ket{1}_{B}$ will create error Bell state detection under the case of lossy channel, which will cause the unbalanced bit value and high bit error rate.

To solve this issue, Alice and Bob need to decrease the probability of qubit $\ket{1}$ preparation and increase the probability of qubit $\ket{0}$ preparation.
Therefore, Alice (Bob) should exploit the entangled state $\ket{\psi}_{t}=\sqrt{1-t}\ket{0}\ket{1}+\sqrt{t}\ket{1}\ket{0}$ to replace the maximally entangled state $\ket{\psi^{+}}=(\ket{0}\ket{1}+\ket{1}\ket{0})/\sqrt{2}$ in the entanglement-based protocol with Fig. \ref{f1}(a), where $t$ is the transmittance of partial BS. Note that the non-maximally entangled state is also used to prove the security in the TF-QKD\cite{curty:2018:simple}.
Taking into account the threshold detector and lossy channel, the joint quantum state of Alice's $a$ mode and Bob's $b$ mode after Charlie's BSM with $\ket{\psi^{\pm}}_{AB}$
 under the case without eavesdropper's disturbance can be written as (see Methods for detail)
\begin{equation}
\begin{aligned}\label{}
\rho_{ab}^{\pm}=\frac{q_{0}}{q}\ket{11}_{ab}\bra{11}+\frac{q_{1}}{q}\ket{\psi^{\pm}}_{ab}\bra{\psi^{\pm}}+\frac{q_{2}}{q}\ket{00}_{ab}\bra{00},
\end{aligned}
\end{equation}
where $q=q_{0}+q_{1}+q_{2}$, $q_{0}$, $q_{1}$ and $q_{2}$ are the probabilities of Charlie's successful BSM given that the photon numbers of TF state are zero, one and two.
Consider a virtual step, if Alice and Bob jointly perform QND measurement on TF state to implement photon-number-resolving before they send TF state to Charlie, the joint quantum state of Alice's $a$ mode and Bob's $b$ mode is $\ket{\psi^{\pm}}_{ab}$ given that the TF state with one-photon and Charlie's BSM with $\ket{\psi^{\pm}}_{AB}$, which reduces to the the case of ideal detector and lossless channel.

Similarly, we can have a equivalent prepare-and-measure protocol corresponding to the entanglement-based protocol with entangled state $\ket{\psi}_{t}=\sqrt{1-t}\ket{0}\ket{1}+\sqrt{t}\ket{1}\ket{0}$.
Alice (Bob) prepares the qubit $\ket{+z}=\ket{0}$ and $\ket{-z}=\ket{1}$ with probability $1-t$ and $t$ as $Z$ basis logic bit 0 and 1, respectively.
Alice (Bob) prepares the qubit $\ket{+x}=\sqrt{1-t}\ket{0}+\sqrt{t}\ket{1}$ and $\ket{-x}=\sqrt{1-t}\ket{0}-\sqrt{t}\ket{1}$ with equal probability as $X$ basis logic bit 0 and 1, respectively.
Alice (Bob) prepares the qubit $\ket{+y}=\sqrt{1-t}\ket{0}+i\sqrt{t}\ket{1}$ and $\ket{-y}=\sqrt{1-t}\ket{0}-i\sqrt{t}\ket{1}$ with equal probability as $Y$ basis logic bit 0 and 1, respectively. Obviously, the quantum state can be seen as a mixture of photon number states for TF state in the $Z$ basis. For the TF state with one-photon in the $Z$ basis, one of Alice and Bob needs to prepare $\ket{0}$ as logic bit 0 and the other prepare $\ket{1}$ as logic bit 1. However, the quantum state is coherent superposition of photon number states for TF state in the $X$ ($Y$) basis. Here, if we assume Alice and Bob knowing the quantum bit error rate (QBER) of TF state with one-photon in the $X$ basis, for example, Alice and Bob can perform joint QND measurement on TF state to implement photon-number-resolving in the $X$ basis, one can use the case of TF state with one-photon to extract secure key in the BB84 encoding, which can be given by (see Methods for detail)
\begin{equation}
\begin{aligned}\label{eq2}
R_{\textrm{BB84}}=q_{1}[1-H(e_{XX}^{b1})]-qH(E_{ZZ}),
\end{aligned}
\end{equation}
where $E_{ZZ}=(q_{0}+q_{2})/q$ is the QBER of $Z$ basis, $H(x)=-x\log_{2}(x)-(1-x)\log_{2}(1-x)$ is the binary Shannon entropy and $e_{XX}^{b1}$ is the QBER in $X$ basis for TF state with one-photon. We can have optimal secure key rate in Eq. \eqref{eq2} with the transmittance of partial BS $t\approx8\%$ given that QBER $e_{XX}^{b1}=3\%$, dark count rate of threshold detector $p_{d}=10^{-6}$, efficiency of threshold detector $\eta_{d}=40\%$ and the fiber distance between Alice and Bob $L\geq 100$ km.
Note that the entanglement-based protocol in Fig. \ref{f1}(a) and prepare-and-measure protocol in Fig. \ref{f1}(b) are the virtual protocols, which are not used to perform experiment but prove the security in theory.

\textbf{TF-QKD with phase-encoding coherent state.}
Manipulating the quantum state with superpositions of the vacuum and one-photon states and, in particular, requiring control about the relative phase between the vacuum and one-photon state is quite problematic\cite{lombardi2002teleportation}. However, we consider the coherent state $\ket{\alpha}=e^{-\mu/2}\sum_{n=0}^{\infty}\frac{(e^{i\theta}\sqrt{\mu})^{n}}{\sqrt{n!}}\ket{n}$, where the relative phase $\theta$ between the different Fock states in the superposition is reflected physically in the phase of the classical electric field. Hereafter, the phase-encoding basis means to implement phase modulation of coherent state, such as $X$ and $Y$ basis.
In order to achieve Alice and Bob knowing the QBER of TF state with one-photon in the phase-encoding basis without the requirement of QND measurement, one can use the post-selected phase-matching method for phase-randomized coherent state\cite{lucamarini2018overcoming,ma2018phase}. By using the post-selected phase-matching method, the phases of Alice's and Bob's coherent state can be seen as equal and randomized, which means that they can use decoy-state method to estimate the yield and QBER of TF state with one-photon in the phase-encoding basis (see Methods).

\emph{Efficient TF-QKD.} Here, we propose an efficient TF-QKD that the single-photon source used for $Z$ basis and laser source used for phase-encoding basis in Fig. \ref{f1}(c). The qubit prepared in $Z$ basis can be implemented by turning on and off (such as optical switch) the single-photon source, while the qubit of phase encoding basis should exploit the phase-randomized coherent state combined with phase modulation. However, the perfect single-photon source is still a challenge under the current technology.
Therefore, we propose a practical TF-QKD by exploiting phase-randomized coherent state to replace single-photon source used for $Z$ basis encoding.

\begin{figure*}
\centering
\includegraphics[width=14cm]{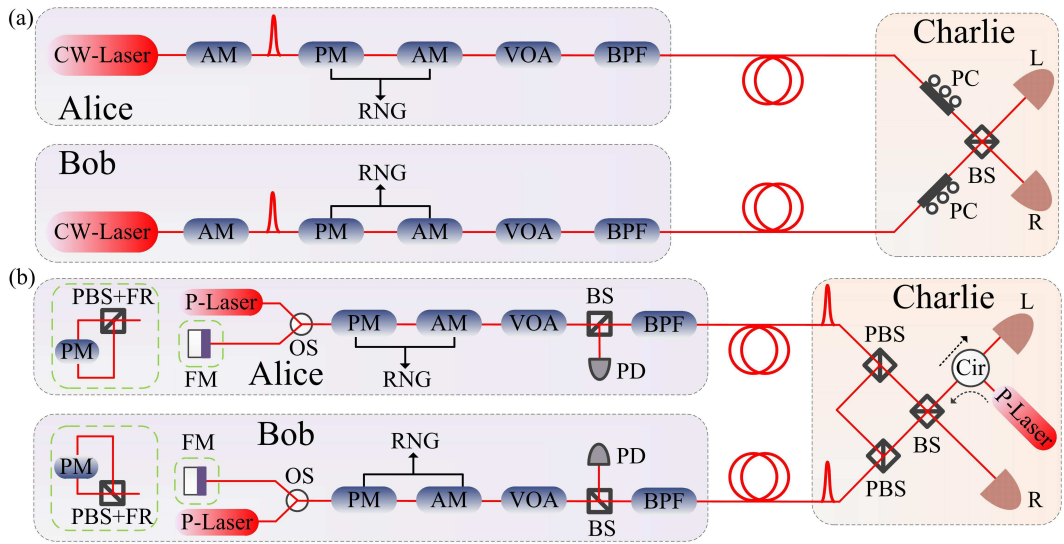}
\caption{The practical TF-QKD setup. (a) practical TF-QKD with independent lasers. The phase modulator (PM) can realize phase encoding and random phase modulation at one time. CW-Laser: continuous-wave laser, AM: amplitude modulator, VOA, variable optical attenuator, BPF: band pass filter, PC: polarization controller, BS: beam splitter, RNG: random number generator.
(b) Phase self-aligned TF-QKD with single laser. The Faraday mirror (FM) or the polarization beam splitter (PBS) and the $\pi/2$ Faraday rotator (FR) are exploited to realize the transformation between horizontal and vertical polarizations. Alice and Bob could choose to prepare the qubit in $Z$ basis by using Charlie's laser or their own pulse lasers. The security will be enhanced if they use their own laser. Some polarization-maintaining fiber are required to keep the polarization in the systems of  Alice, Bob and Charlie. P-Laser: pulse laser, OS: optical switch, PD, photoelectric detector, Cir: circulator.
} \label{f2}
\end{figure*}

\emph{Practical TF-QKD.}
In the following, let us explain our practical TF-QKD in detail as shown in Fig. \ref{f2}(a). (i) Alice and Bob use the stabilized narrow line-width continuous-wave laser and amplitude modulator to prepare the global phase stabilized optical pulses. Alice's and Bob's random phases $\theta_{A}\in [0,2\pi)$ and $\theta_{B}\in [0,2\pi)$ are realized by using phase modulators. For $Z$ basis encoding, the phase-randomized coherent state with intensities $0$ and $\mu$ as logic bits 0 and 1 with probabilities $1-t$ and $t$ by using amplitude modulator. For $X$ ($Y$) basis encoding, they use the phase and amplitude modulator to randomly implement $0$ ($\pi/2$) and $\pi$ ($-\pi/2$) phase modulation as logic bits 0 and 1 with intensities $\{\nu/2, \omega/2, 0\}$.
(ii) Then they send quantum states to Charlie for single-photon BSM through the insecure quantum channel. Charlie publishes the successful events of single-photon BSM. (iii) Alice and Bob will announce the basis information through the authenticated classical channel. The intensity and random phase information $k_{A,B}$ of phase-encoding basis should be disclosed, while those of $Z$ basis are confidential to Charlie, where they have $\theta_{A,B}\in \Delta_{k_{A,B}}$, $\Delta_{k_{A,B}}=[\frac{2\pi k_{A,B}}{M},\frac{2\pi (k_{A,B}+1)}{M})$ and $k_{A,B}\in\{0,1,\ldots,M-1\}$. (iv) Alice and Bob use the data of $Z$ basis as the raw key, while the data of phase-encoding basis are announced to estimate the amount of leaked information. (v) They exploit the classical error correction and privacy amplification to extract the secure key rate.

After Charlie announces the measurement results, he cannot change the yield and QBER due to information causality\cite{pawlowski2009information}. The decoy-state method of estimating the yield and QBER of TF state with $n$-photon in phase-encoding basis is also true even for the post-selected phase-matching method, which has also been used in phase-matching QKD\cite{ma2018phase}.
The GLLP analysis\cite{gottesman2004Security} can be used for the data of $Z$ basis, since the random phases information of Alice's and Bob's coherent states are all confidential to Charlie.
Bob will always flit his bit in $Z$ basis. Due to the density matrix of TF state with one-photon $\rho_{\textrm{TF}}^{1ZZ}=\rho_{\textrm{TF}}^{1XX}=\frac{1}{2}(\ket{01}_{AB}\bra{01}+\ket{10}_{AB}\bra{10})$, we can use the yield of TF state with one-photon $Y_{\textrm{TF}}^{1ZZ}=Y_{\textrm{TF}}^{1XX}$ in the asymptotic limit.
Note that, we can also directly estimate the yield $Y_{\textrm{TF}}^{1ZZ}$ by using the data of phase-encoding basis given that one of Alice and Bob sends intensity $0$.

For the BB84 encoding\cite{bennett1984quantum}, Alice and Bob only keep the data of $|k_{B}-k_{A}|=0$ and $M/2$ when they both choose $X$ basis by the post-selected phase-matching method. If $|k_{B}-k_{A}|=0$ $\left(|k_{B}-k_{A}|=M/2\right)$, Bob will flit his bit when Charlie receives a Bell state $\ket{\psi^{-}}_{AB}$ $\left(\ket{\psi^{+}}_{AB}\right)$.
The secure key rate of practical TF-QKD can be given by
\begin{equation}
\begin{aligned}\label{eq1}
R_{\textrm{TF-BB84}}=2t(1-t)\mu e^{-\mu}Y_{\textrm{TF}}^{1ZZ}[1-H(e_{XX}^{b1})]-Q_{ZZ}fH(E_{ZZ}),
\end{aligned}
\end{equation}
where $Q_{ZZ}$ is the gain in $Z$ basis acquired directly from the experiment, $f=1.15$ is the error correction coefficient.

For the RFI scheme\cite{laing2010reference,wang2015phase}, one can allow Alice and Bob to have different phase references which can be changed slowly (details can be found in Methods). Therefore, they can collect the data of $|k_{B}-k_{A}|=k$, $k\in\{0,1,\ldots,M-1\}$ to form a set $D_{k}$, where the probability of $|k_{B}-k_{A}|=k$ is $\frac{1}{M}$. For each set $D_{k}$, they calculate the value $C_{k}^{1}=(1-2e_{XXk}^{b1})^2+(1-2e_{XYk}^{b1})^2+(1-2e_{YXk}^{b1})^2+(1-2e_{YYk}^{b1})^2$, where $e_{XXk(XYk,YXk,YYk)}^{b1}$ is the QBER of TF state with one-photon in set $D_{k}$ given that Alice and Bob choose $X-X(X-Y,Y-X,Y-Y)$ basis.
The secure key rate of practical TF-QKD with RFI scheme can be given by
\begin{equation}
\begin{aligned}\label{eq3}
R_{\textrm{TF-RIF}}=&2t(1-t)\mu e^{-\mu}Y_{\textrm{TF}}^{1ZZ}[1-I_{E}(C^{1})]-Q_{ZZ}fH(E_{ZZ}),
\end{aligned}
\end{equation}
where $I_{E}(C^{1})=(1-e_{ZZ}^{b1})H(\frac{1+\mu}{2})+e_{ZZ}^{b1}H(\frac{1+v}{2})$ describes eavesdropper Eve's information, thereinto, $v=\sqrt{C^{1}/2-(1-e_{ZZ}^{b1})^2u^2}/e_{ZZ}^{b1}$, $u=\textrm{min}[\sqrt{C^{1}/2}/(1-e_{ZZ}^{b1}),1]$ and $C^{1}=\frac{1}{M}\sum_{k=0}^{M-1}C_{k}^{1}$.
Compared with the BB84 encoding, all data of RFI scheme can be exploited to estimate parameter $C^{1}$, which can be used to slow down the finite size effect. Alice and Bob can change $M$ to acquire the maximum key rate without impacts on efficiency.  The QBER of $Z$ basis for TF state with one-photon $e_{ZZ}^{b1}\equiv0$ leads to $I_{E}(C^{1})=H((1+\sqrt{C^{1}/2})/2)$.

\begin{figure*}
\centering
\includegraphics[width=14cm]{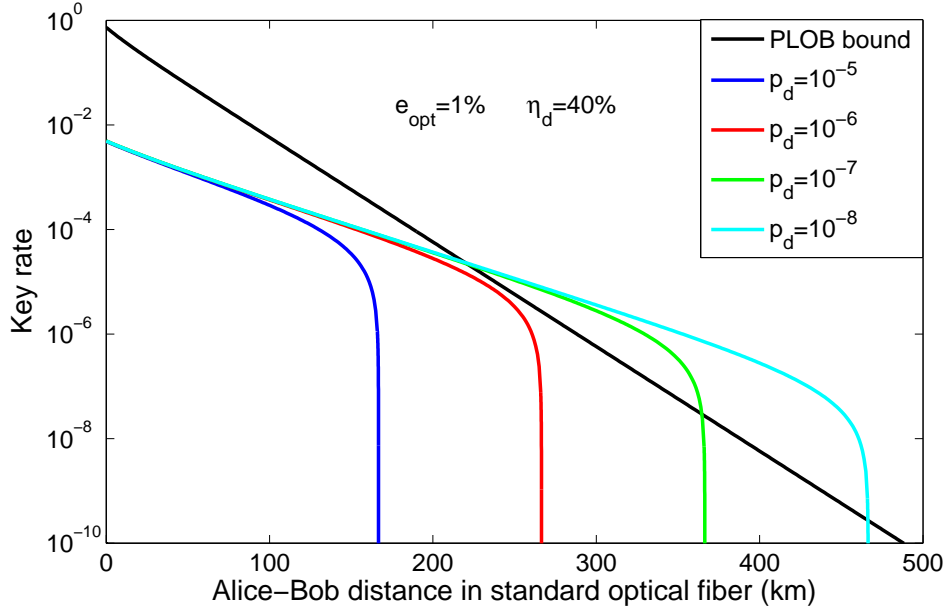}
\caption{The key rate of practical TF-QKD with BB84 encoding in the asymptotic limit. For each transmission loss, we
optimize the parameters $\mu$ and $t$ with $e_{\textrm{opt}}=1\%$, $\nu=0.1$, $\omega=0.02$ and $M=16$. For the PLOB bound, we use $R_{\textrm{PLOB}}=-\log_{2}(1-\eta_{\textrm{PLOB}})$, $\eta_{\textrm{PLOB}}=\eta_{d}\times10^{-0.02L}$. The secure key rate of TF-QKD with BB84 encoding can surpass the PLOB bound under the case of detector with $\eta_{d}=40\%$, $p_{d}=10^{-7}$, the performance of detector has been realized much more\cite{Yin:2016:Measurement}.
} \label{f3}
\end{figure*}

\begin{figure*}
\centering
\includegraphics[width=14cm]{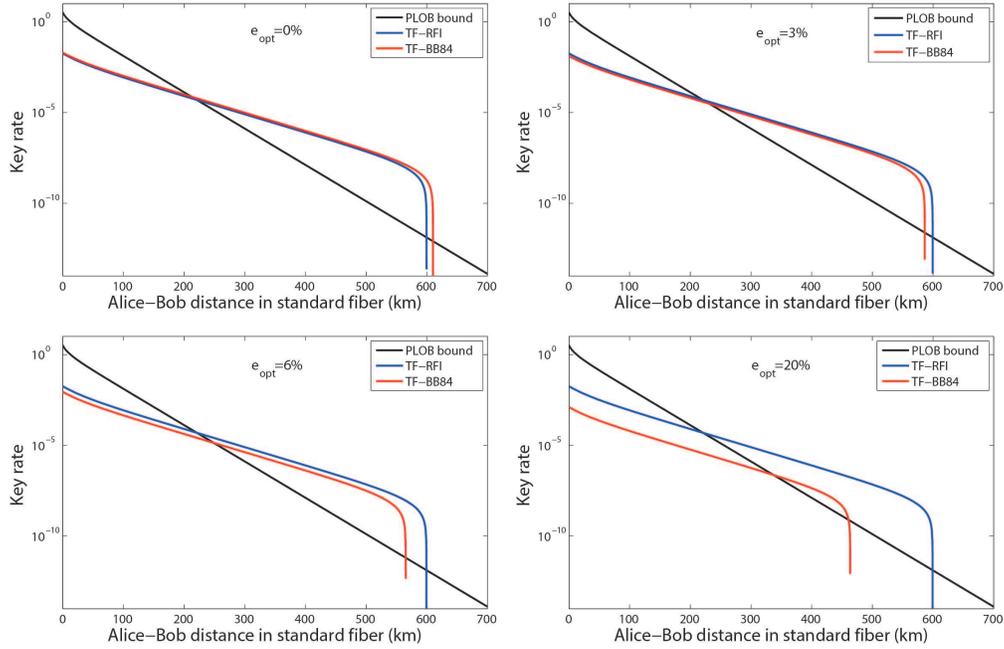}
\caption{The key rates of practical TF-QKD with BB84 encoding and RFI scheme in the asymptotic limit. For each transmission loss, we
optimize the parameters $\mu$ and $t$ with $\eta_{d}=90\%$, $p_{d}=10^{-9}$, $\nu=0.1$, $\omega=0.02$ and $M=16$. The secure key rate of practical TF-QKD with RFI scheme do not change obviously with optical error rate $e_{\textrm{opt}}$. The secure key rate of practical TF-QKD with BB84 encoding can also beat the PLOB bound even the optical error rate up to $e_{\textrm{opt}}=20\%$.
} \label{f4}
\end{figure*}

The secure key rate of practical TF-QKD using BB84 encoding changes with the dark count rate as shown in Fig. \ref{f3}. We use the practical parameters to simulate the secure key rate in Fig. \ref{f3}, where the efficiency of detector is $\eta_{d}=40\%$, the loss coefficient of the channel is $0.2$ dB/km and the optical error rate of system is $e_{\textrm{opt}}=1\%$.
The optical error rate is usually large due to the long-distance single-photon-type interference. We compare the secure key rates of practical TF-QKD using BB84 encoding and RFI scheme with the different optical error rate as shown in Fig. \ref{f4}.
To show the advantage of TF-QKD, the efficiency and dark count rate of detector are assumed to be $\eta_{d}=90\%$ and $p_{d}=10^{-9}$ in Fig. \ref{f4}, respectively. In the simulation, both schemes can surpass the PLOB bound and tolerate the big optical error rate $e_{\textrm{opt}}$. The key rate of TF-QKD with BB84 encoding will significantly decline with $e_{\textrm{opt}}$ rising, while the RFI scheme is robust.
However, the long-distance phase-stabilization (it could not be a perfect match but is required to vary slowly) also exists since the relative phase changes too fast in the long-distance fiber or free-space channel.

The experimental demonstration of TF-QKD with independent lasers in Fig. \ref{f2}(a) is a big challenge, although the MDI-QKD with two-photon BSM has been implemented over 404 km optical fiber\cite{Yin:2016:Measurement} by using asymmetric four-intensity decoy-state method\cite{Zhou:2016:Making}. Compared with the two-photon BSM, greater technological challenges exist in the TF-QKD with single-photon BSM. The frequency difference of two independent lasers is required more rigorously\cite{lucamarini2018overcoming}. The phase-locking technique may be used to compensate the frequency difference. Importantly, the long-distance phase-stabilization technique is required to implement single-photon interference with phase matching. The RFI scheme can allow the phase mismatching. However, the relative phase change is still required to vary slowly. To rapidly implement the proof-of-principle TF-QKD experiment, we present a phase self-aligned TF-QKD with single laser interference as shown in Fig. \ref{f2}(b). The horizontal polarization optical pulse generated by Charlie is divided into two pulses by the polarization-maintaining beam splitter. By exploiting the $\pi/2$ rotation effect of Faraday mirror, the two pulses interfere after they go through the same path. Though the phase self-aligned scheme would be affected by the loss and noise, the frequency difference and long-distance phase-stabilization problems are both solved\cite{guan2016observation}. An extra security analysis
with untrusted source\cite{zhao2010security} should be used to defeat the attack from systems of Alice and Bob.

\section*{Discussion}

In summary, we have proved that the TF-QKD can be regarded as a MDI-QKD with single-photon BSM.
By introducing the $Z$ basis encoding, the secret key extraction can exploit the tagging method of GLLP analysis and the decoy-state method.
Compared with BB84 encoding, the RFI scheme has the advantages of increasing the data of parameter estimation and reducing the effect of phase drift. We should point out that the extra $Y$ basis preparation in RFI scheme does not add additional operation due to the active phase randomization requirement, which is different from the traditional QKD. We propose a feasible experimental scheme to implement the proof-of-principle experimental demonstration. Note that, the security of this proof-of-principle experiment in Fig. \ref{f2}b is not guaranteed with our current analysis, which requires a further security evaluation due to introducing untrusted source. Through simulation, we show that the secure key rate of practical TF-QKD can surpass the PLOB bound. The universally composable security with finite-key analysis needs to be considered in the future. Our proposal suggests an important avenue for practical high-speed and long-distance QKD without detector vulnerabilities. During the preparation of this paper and posting it on the arXiv, we became aware of some important works\cite{tamaki2018information,ma2018phase,Wang2018Sending,cui:2018:phase,curty:2018:simple,Lin:2018:A} of TF-QKD.

\section*{Methods}

\textbf{MDI-QKD with single-photon BSM.}
For the case of entanglement-based protocol with the entangled state $\ket{\psi}_{t}=\sqrt{1-t}\ket{0}\ket{1}+\sqrt{t}\ket{1}\ket{0}$, the joint quantum state of Alice and Bob can be given by
\begin{equation}
\begin{aligned}\label{}
\ket{\varphi}_{ABab}&=\ket{\psi^{+}}_{Aa}\otimes\ket{\psi^{+}}_{Bb}\\
&=(1-t)\ket{0011}_{ABab}+\sqrt{t(1-t)}\left(\ket{0110}_{ABab}+\ket{1001}_{ABab}\right)+t\ket{1100}_{ABab}.
\end{aligned}
\end{equation}
For the threshold detector and lossy channel, the TF state $\ket{00}_{AB}$, $\ket{01}_{AB}$, $\ket{10}_{AB}$ and $\ket{11}_{AB}$  will all have single-photon Bell state clicks. Due to the single-photon BSM of Charlie, the photon number of TF state will collapse to three events, namely vacuum, one-photon and two-photon. The corresponding probability can be expressed as
\begin{equation}
\begin{aligned}\label{}
q_{0}&=2(1-t)^{2}p_{d}(1-p_{d}),\\
q_{1}&=2t(1-t)\left\{p_{d}(1-p_{d})(1-\sqrt{\eta})+(1-p_{d})[1-(1-p_{d})(1-\sqrt{\eta})]\right\},\\
q_{2}&=t^{2}\left\{p_{d}(1-p_{d})(1-\sqrt{\eta})^2+(1-p_{d})[1-(1-p_{d})(1-\sqrt{\eta})^2]\right\},
\end{aligned}
\end{equation}
where the expression of $q_{2}$ is acquired by the Hong-Ou-Mandel interference of two-photon. The parameter $\sqrt{\eta}=\eta_{d}\times10^{-0.02L/2}$ is the transmittance between Alice (Bob) and Charlie.

For the case of prepare-and-measure protocol corresponding to entanglement-based protocol with the entangled state $\ket{\psi}_{t}=\sqrt{1-t}\ket{0}\ket{1}+\sqrt{t}\ket{1}\ket{0}$, the density matrix of TF state in the $Z$ basis is
\begin{equation}
\begin{aligned}\label{}
\rho_{\textrm{TF}}^{ZZ}=t(1-t)(\ket{01}_{AB}\bra{01}+\ket{10}_{AB}\bra{10})+(1-t)^2\ket{00}_{AB}\bra{00}+t^2\ket{11}_{AB}\bra{11},
\end{aligned}
\end{equation}
which means a mixture of photon number states for TF state in the $Z$ basis. The TF state of $Z$ basis is the product state of Alice's and Bob's quantum state. The density matrix of TF state with one-photon in the $Z$ basis is
\begin{equation}
\begin{aligned}\label{}
\rho_{\textrm{TF}}^{1ZZ}=\frac{1}{2}(\ket{01}_{AB}\bra{01}+\ket{10}_{AB}\bra{10}),
\end{aligned}
\end{equation}
which needs one of Alice and Bob prepares $\ket{0}$ as logic bit 0 and the other prepares $\ket{1}$ as logic bit 1.

The density matrix of TF state in the $X$ basis can be written as
\begin{equation}
\begin{aligned}\label{}
\rho_{\textrm{TF}}^{XX}=&\frac{1}{4}\big[\ket{+x,+x}_{AB}\bra{+x,+x}+\ket{+x,-x}_{AB}\bra{+x,-x}\\
&+\ket{-x,+x}_{AB}\bra{-x,+x}+\ket{-x,-x}_{AB}\bra{-x,-x}\big].
\end{aligned}
\end{equation}
Thereinto, we have
\begin{equation}
\begin{aligned}\label{}
\ket{\pm x,+x}_{AB}&=(1-t)\ket{00}_{AB}+\sqrt{t(1-t)}(\ket{01}_{AB}\pm\ket{10}_{AB})\pm t\ket{11}_{AB},\\
\ket{\pm x,-x}_{AB}&=(1-t)\ket{00}_{AB}-\sqrt{t(1-t)}(\ket{01}_{AB}\mp\ket{10}_{AB})\mp t\ket{11}_{AB},\\
\end{aligned}
\end{equation}
which means a coherent superposition of photon number state for TF state in the $X$ basis.
If Alice and Bob jointly perform QND measurement on TF state to implement photon-number-resolving, we have
\begin{equation}
\begin{aligned}\label{}
\ket{\pm x,+x}_{AB}\autorightarrow{QND measurement}{one-photon}\frac{1}{\sqrt{2}}(\ket{01}_{AB}\pm\ket{10}_{AB}),\\
\ket{\pm x,-x}_{AB}\autorightarrow{QND measurement}{one-photon}\frac{1}{\sqrt{2}}(\ket{01}_{AB}\mp\ket{10}_{AB}),\\
\rho_{\textrm{TF}}^{1XX}=\frac{1}{2}(\ket{01}_{AB}\bra{01}+\ket{10}_{AB}\bra{10})=\rho_{\textrm{TF}}^{1Z},
\end{aligned}
\end{equation}
where $\rho_{\textrm{TF}}^{1ZZ}$ ($\rho_{\textrm{TF}}^{1XX}$) is the density matrix of TF state with one-photon in the $Z$ ($X$) basis.
We have $Y_{\textrm{TF}}^{1ZZ}=Y_{\textrm{TF}}^{1XX}$ in the asymptotic limit due to $\rho_{\textrm{TF}}^{1ZZ}=\rho_{\textrm{TF}}^{1XX}$, where $Y_{\textrm{TF}}^{1ZZ}$ ($Y_{\textrm{TF}}^{1XX}$) is the yield given that Alice and Bob choose $Z$ ($X$) basis and TF state contains one-photon.
Alice and Bob can know the locations of the TF state with one-photon by using the QND measurement, they could discard all other states and apply error correction and privacy amplification only to the TF state with one-photon. In this case with BB84 encoding, they can achieve a key rate of\cite{lo1999unconditional,shor2000simple}
\begin{equation}
\begin{aligned}\label{}
R_{\textrm{BB84}}=q_{1}[1-H(e_{ZZ}^{b1})-H(e_{XX}^{b1})]).
\end{aligned}
\end{equation}
For the TF state with one-photon in the $Z$ basis, we have $e_{ZZ}^{b1}\equiv0$ since we only have the case of Alice's logic bit 0 (1) and Bob's logic bit 1 (0) corresponding to quantum state $\ket{01}$ \big($\ket{10}$\big).

However, if we assume that Alice and Bob can know the QBER of TF state with one-photon in the $X$ basis, one can acquire the secure key in the $Z$ basis without Alice and Bob
knowing the locations (QND measurement) of the TF state with one-photon by using the GLLP analysis\cite{gottesman2004Security}. The secure key rate can be given by
\begin{equation}
\begin{aligned}\label{}
R_{\textrm{BB84}}=q_{1}[1-H(e_{XX}^{b1})]-qH(E_{ZZ}),
\end{aligned}
\end{equation}
where the parameter $q_{1}$ should be calculated by using the decoy-state method, for example, we choose three value of $t$ in the $Z$ basis.

\textbf{TF-QKD with phase-encoding coherent state.}
In order to make Alice and Bob know the QBER of TF state with one-photon in the $X$ basis without the requirement of QND measurement, we need to consider the case of phase-randomized coherent state
\begin{equation}
\begin{aligned}\label{}
\rho &=\frac{1}{2\pi}\int_{0}^{2\pi}\ket{\alpha}_{A}\bra{\alpha}\otimes\ket{e^{i\delta}\alpha}_{B}\bra{e^{i\delta}\alpha}d\theta\\
&=e^{-2\mu}\sum_{n=0}^{\infty}\sum_{m=0}^{\infty}\sum_{k=0}^{n+m}\frac{e^{i\delta(k-n)}\mu^{n+m}}{\sqrt{n!m!k!(n+m-k)!}}\ket{n}_{A}\bra{k}\otimes\ket{m}_{B}\bra{n+m-k},
\end{aligned}
\end{equation}
where the global phases of Alice's coherent state $\ket{\alpha}_{A}=\ket{e^{i\theta}\sqrt{\mu}}_{A}$ and Bob's $\ket{e^{i\delta}\alpha}_{B}=\ket{e^{i(\theta+\delta)}\sqrt{\mu}}_{B}$ should be randomized and have a fixed phase difference $\delta$. Therefore, we have
\begin{equation}
\begin{aligned}\label{}
&\ket{\alpha}_{A}\ket{e^{i\delta}\alpha}_{B}\autorightarrow{phase-randomized}{one-photon}\frac{1}{\sqrt{2}}(\ket{01}_{AB}+e^{-i\delta}\ket{10}_{AB}).\\
\end{aligned}
\end{equation}
For the $X$ basis encoding, we have
\begin{equation}
\begin{aligned}\label{}
&\ket{\pm \alpha}_{A}\ket{+\alpha}_{B}\autorightarrow{phase-randomized}{one-photon}\frac{1}{\sqrt{2}}(\ket{01}_{AB}\pm\ket{10}_{AB}),\\
&\ket{\pm \alpha}_{A}\ket{-\alpha}_{B}\autorightarrow{phase-randomized}{one-photon}\frac{1}{\sqrt{2}}(\ket{01}_{AB}\mp\ket{10}_{AB}),\\
&\rho_{\textrm{TF}}^{1XX}=\frac{1}{2}(\ket{01}_{AB}\bra{01}+\ket{10}_{AB}\bra{10}),
\end{aligned}
\end{equation}
where the global phases of Alice's and Bob's coherent state should be equal and randomized. It can be realized by using post-selected phase-matching method for phase-randomized coherent state introduced in the original TF-QKD\cite{lucamarini2018overcoming} and phase-matching QKD\cite{ma2018phase}. If we consider the photon number space of TF state given that the global phases of Alice's coherent state and Bob's are randomized and have a fixed phase difference, the density matrix can be given by
\begin{equation}
\begin{aligned}\label{}
\rho=e^{-2\mu}\sum_{n=0}^{\infty}\frac{(2\mu)^{n}}{n!}\ket{n}_{\textrm{TF}}\bra{n},
\end{aligned}
\end{equation}
which is similar with the phase encoding phase-randomized coherent state in the traditional decoy-state QKD\cite{wang2005beating,lo2005decoy}.
Therefore, the decoy state method can be used for estimating the yield and QBER of TF state with one-photon.

For phase-randomized coherent state used for $Z$ basis encoding, we have
\begin{equation}
\begin{aligned}\label{}
\rho_{\textrm{TF}}^{ZZ}=&(1-t)^2\ket{00}_{AB}\bra{00}+t^2\left(\sum_{n=0}^{\infty}e^{-\mu}\frac{\mu^{n}}{n!}\ket{n}_{A}\bra{n}\right)\left(\sum_{m=0}^{\infty}e^{-\mu}\frac{\mu^{m}}{m!}\ket{m}_{B}\bra{m}\right)\\
&+t(1-t)\left[\ket{0}_{A}\bra{0}\left(\sum_{n=0}^{\infty}e^{-\mu}\frac{\mu^{n}}{n!}\ket{n}_{B}\bra{n}\right)+\left(\sum_{n=0}^{\infty}e^{-\mu}\frac{\mu^{n}}{n!}\ket{n}_{A}\bra{n}\right)\ket{0}_{B}\bra{0}\right].
\end{aligned}
\end{equation}
We need $\ket{0}$ as logic bit 0 and $\ket{1}$ as logic bit 1, therefore the efficient TF state with one-photon in $Z$ basis only results from the case of logic bit $0_{A}1_{B}$ and $1_{A}0_{B}$ with the probability $2t(1-t)\mu e^{-\mu}$.
For simulation, we consider the case without Charlie's disturbance. In the $Z$ basis of practical TF-QKD, by going through the quantum channel and beam splitter, we have $(1-t)^{2}$ probability of quantum state
\begin{equation}
\begin{aligned}\label{}
\ket{0}_{A}\ket{0}_{B}\xrightarrow{\textrm{BS}}\ket{0}_{L}\ket{0}_{R},
\end{aligned}
\end{equation}
$t(1-t)$ probability of quantum state
\begin{equation}
\begin{aligned}\label{}
\ket{0}_{A}\ket{e^{i\theta_{B}}\sqrt{\mu}}_{B}\xrightarrow{\textrm{BS}}\left|e^{i\theta_{B}}\sqrt{\frac{\mu\sqrt{\eta}}{2}}\right\rangle_{L}\left|-e^{i\theta_{B}}\sqrt{\frac{\mu\sqrt{\eta}}{2}}\right\rangle_{R},
\end{aligned}
\end{equation}
$t(1-t)$ probability of quantum state
\begin{equation}
\begin{aligned}\label{}
\ket{e^{i\theta_{A}}\sqrt{\mu}}_{A}\ket{0}_{B}\xrightarrow{\textrm{BS}}\left|e^{i\theta_{A}}\sqrt{\frac{\mu\sqrt{\eta}}{2}}\right\rangle_{L}\left|e^{i\theta_{A}}\sqrt{\frac{\mu\sqrt{\eta}}{2}}\right\rangle_{R},
\end{aligned}
\end{equation}
and $t^2$probability of quantum state
\begin{equation}
\begin{aligned}\label{}
\ket{e^{i\theta_{A}}\sqrt{\mu}}_{A}\ket{e^{i\theta_{B}}\sqrt{\mu}}_{B}\xrightarrow{\textrm{BS}}\left|e^{i\theta_{A}}\sqrt{\frac{\mu\sqrt{\eta}}{2}}+e^{i\theta_{B}}\sqrt{\frac{\mu\sqrt{\eta}}{2}}\right\rangle_{L}\left|e^{i\theta_{A}}\sqrt{\frac{\mu\sqrt{\eta}}{2}}-e^{i\theta_{B}}\sqrt{\frac{\mu\sqrt{\eta}}{2}}\right\rangle_{R}.
\end{aligned}
\end{equation}
Here, we have $\theta_{A}\in [0,2\pi)$ and $\theta_{B}\in [0,2\pi)$, $L$ and $R$ represent the left detector and right detector of Charlie, respectively.
The gain $Q_{ZZ}$ and QBER $E_{ZZ}$ of practical TF-QKD can be given by
\begin{equation}
\begin{aligned}\label{}
Q_{ZZ}=&2p_{d}(1-p_{d})(1-t)^2+4(1-p_{d})e^{-\frac{\mu}{2}\sqrt{\eta}}[1-(1-p_{d})e^{-\frac{\mu}{2}\sqrt{\eta}}]t(1-t)\\
&+2(1-p_{d})e^{-\mu\sqrt{\eta}}[I_{0}(\mu\sqrt{\eta})-(1-p_{d})e^{-\mu\sqrt{\eta}}]t^2,
\end{aligned}
\end{equation}
and
\begin{equation}
\begin{aligned}\label{}
E_{ZZ}Q_{ZZ}=&2p_{d}(1-p_{d})(1-t)^2+2(1-p_{d})e^{-\mu\sqrt{\eta}}[I_{0}(\mu\sqrt{\eta})-(1-p_{d})e^{-\mu\sqrt{\eta}}]t^2.
\end{aligned}
\end{equation}

For phase-encoding basis of practical TF-QKD, by going through the quantum channel and beam splitter, we have $1/4$ probability of quantum state
\begin{equation}
\begin{aligned}\label{}
&\ket{e^{i(\theta_{A}+\pi g_{A}+\frac{\pi}{2}h_{A})}\sqrt{\lambda}}_{A}\ket{e^{i(\theta_{B}+\pi g_{B}+\frac{\pi}{2}h_{B})}\sqrt{\chi}}_{B}\\
&\xrightarrow{\textrm{BS}}\left|e^{i(\theta_{A}+\pi g_{A}+\frac{\pi}{2}h_{A})}\sqrt{\frac{\lambda\sqrt{\eta}}{2}}+e^{i(\theta_{B}+\pi g_{B}+\frac{\pi}{2}h_{B})}\sqrt{\frac{\chi\sqrt{\eta}}{2}}\right\rangle_{L}\\
&\otimes\left|e^{i(\theta_{A}+\pi g_{A}+\frac{\pi}{2}h_{A})}\sqrt{\frac{\lambda\sqrt{\eta}}{2}}-e^{i(\theta_{B}+\pi g_{B}+\frac{\pi}{2}h_{B})}\sqrt{\frac{\chi\sqrt{\eta}}{2}}\right\rangle_{R},
\end{aligned}
\end{equation}
where $h_{A},h_{B}\in\{0,1\}$ represent basis $X$ and $Y$, $g_{A},g_{B}\in\{0,1\}$ represent logic bit 0 and 1 given that the intensities of Alice's and Bob's are $\lambda$ and $\chi$, respectively, $\lambda,\chi\in\{\nu/2, \omega/2, 0\}$.
Here, we define $Q^{\theta_{A},\theta_{B},\lambda,\chi}_{h_{A},h_{B}}$ and $E^{\theta_{A},\theta_{B},\lambda,\chi}_{h_{A},h_{B}}$ are the gain and QBER that Alice and Bob choose basis $h_{A}$ and $h_{B}$ when they send the global phase $\theta_{A}$ and $\theta_{B}$ optical pulses with intensities $\lambda$ and $\chi$, respectively. Here,
\begin{equation}
\begin{aligned}\label{}
Q^{\theta_{A},\theta_{B},\lambda,\chi}_{h_{A},h_{B}}=&(1-p_{d})e^{-\frac{\lambda+\chi}{2}\sqrt{\eta}}\left[e^{-\sqrt{\lambda\chi\eta}\cos x}
+e^{\sqrt{\lambda\chi\eta}\cos x}\right]-2(1-p_{d})^{2}e^{-\left(\lambda+\chi\right)\sqrt{\eta}},
\end{aligned}
\end{equation}
and
\begin{equation}
\begin{aligned}\label{}
E^{\theta_{A},\theta_{B},\lambda,\chi}_{h_{A},h_{B}}Q^{\theta_{A},\theta_{B},\lambda,\chi}_{h_{A},h_{B}}=(1-p_{d})e^{-\left(\frac{\lambda+\chi}{2}+\sqrt{\lambda\chi}\cos x\right)\sqrt{\eta}}-(1-p_{d})^{2}e^{-\left(\lambda+\chi\right)\sqrt{\eta}},
\end{aligned}
\end{equation}
where $x=\theta_{B}-\theta_{A}+\frac{\pi}{2}(h_{B}-h_{A})$, $E_{h_{A},h_{B}}^{\theta_{A},\theta_{B},\lambda,\lambda}\simeq \frac{1-\cos x}{2}$ when we assume $\sqrt{\eta}\rightarrow0$ and $p_{d}\rightarrow0$.

Obviously, we can directly estimate the yield $Y_{\textrm{TF}}^{1ZZ}$ by using the data of phase-encoding basis given that one of Alice and Bob sends intensity $0$.
We define $\lambda\uplus\chi$ as the intensity set when Alice and Bob send intensity $\lambda$ and $\chi$ phase-randomized coherent state. Therefore,
$Q^{\frac{\nu}{2}}$, $Q^{\frac{\nu}{2}}$ and $Q^{0}$ are the gain when Alice and Bob send intensities set $\{0\uplus\frac{\nu}{2},\frac{\nu}{2}\uplus0\}$, $\{0\uplus\frac{\omega}{2},\frac{\omega}{2}\uplus0\}$ and $\{0\uplus0\}$, which can be written as
\begin{equation}
\begin{aligned}\label{}
Q^{\frac{\nu}{2}}&=\frac{1}{2}\left(\frac{1}{4\pi^2}\int_{0}^{2\pi}\int_{0}^{2\pi}Q^{\theta_{A},\theta_{B},0,\frac{\nu}{2}}_{h_{A},h_{B}}d\theta_{A}d\theta_{B}+\frac{1}{4\pi^2}\int_{0}^{2\pi}\int_{0}^{2\pi}Q^{\theta_{A},\theta_{B},\frac{\nu}{2},0}_{h_{A},h_{B}}d\theta_{A}d\theta_{B}\right)\\
&=2(1-p_{d})e^{-\frac{\nu}{4}\sqrt{\eta}}\left[1-(1-p_{d})e^{-\frac{\nu}{4}\sqrt{\eta}}\right],\\
Q^{\frac{\omega}{2}}&=2(1-p_{d})e^{-\frac{\omega}{4}\sqrt{\eta}}\left[1-(1-p_{d})e^{-\frac{\omega}{4}\sqrt{\eta}}\right],\\
Q^{0}&=2p_{d}(1-p_{d}).
\end{aligned}
\end{equation}
The $Y_{\textrm{TF}}^{0ZZ}$ and $Y_{\textrm{TF}}^{1ZZ}$ are the yields of TF state with vacuum and one-photon in the $Z$ basis, respectively, which can be given by ($\nu>\omega>0$)\cite{wang2005beating,lo2005decoy}
\begin{equation}
\begin{aligned}\label{}
Y_{\textrm{TF}}^{0ZZ}=Y_{0}=Q^{0}=2p_{d}(1-p_{d}),
\end{aligned}
\end{equation}
and
\begin{equation}
\begin{aligned}\label{}
Y_{\textrm{TF}}^{1ZZ}\geq Y_{\textrm{TF}}^{1ZZL}=\frac{2\nu}{\nu  \omega -\omega ^2} \left(e^{\frac{\omega}{2} } Q^{\frac{\omega}{2}}-\frac{ \omega ^2  }{\nu ^2}e^{\frac{\nu}{2} }Q^{\frac{\nu}{2}}-\frac{\nu ^2-\omega ^2}{\nu ^2}Q^{0}\right).\\
\end{aligned}
\end{equation}

We assume that the optical error rate $e_{\textrm{opt}}$ of $X$ basis exists due to the single-photon interference. For simplicity, we assume that the optical error rate is introduced by the phase misalignment\cite{lucamarini2018overcoming}. Here, a fixed phase difference between Alice's and Bob's global phase is $\delta_{0}=\arccos(1-2e_{\textrm{opt}})$.
By using the post-selected phase-matching method in practical TF-QKD with BB84 encoding, $Q_{XX}^{\nu}$ $\left(Q_{XX}^{\omega}\right)$  and $E_{XX}^{\nu}$ $\left(E_{XX}^{\omega}\right)$ are gain and QBER given that Alice chooses $X$ basis with intensity $\frac{\nu}{2}$ $\left(\frac{\omega}{2}\right)$ and Bob chooses $X$ basis with intensity $\frac{\nu}{2}$ $\left(\frac{\omega}{2}\right)$ in the case of $|k_{B}-k_{A}|=0$ and $\frac{M}{2}$. They can be given by
\begin{equation}
\begin{aligned}\label{}
Q_{XX}^{\nu}&=\frac{M^2}{4\pi^2}\int_{\delta_{0}}^{\delta_{0}+\frac{2\pi}{M}}\int_{0}^{\frac{2\pi}{M}}Q^{\theta_{A},\theta_{B},\frac{\nu}{2},\frac{\nu}{2}}_{0,0}d\theta_{A}d\theta_{B},
~~Q_{XX}^{\omega}&=\frac{M^2}{4\pi^2}\int_{\delta_{0}}^{\delta_{0}+\frac{2\pi}{M}}\int_{0}^{\frac{2\pi}{M}}Q^{\theta_{A},\theta_{B},\frac{\omega}{2},\frac{\omega}{2}}_{0,0}d\theta_{A}d\theta_{B},\\
\end{aligned}
\end{equation}
and
\begin{equation}
\begin{aligned}\label{}
E_{XX}^{\nu}Q_{XX}^{\nu}&=\frac{M^2}{4\pi^2}\int_{\delta_{0}}^{\delta_{0}+\frac{2\pi}{M}}\int_{0}^{\frac{2\pi}{M}}E^{\theta_{A},\theta_{B},\frac{\nu}{2},\frac{\nu}{2}}_{0,0}Q^{\theta_{A},\theta_{B},\frac{\nu}{2},\frac{\nu}{2}}_{0,0}d\theta_{A}d\theta_{B},\\
Q_{XX}^{\omega}Q_{XX}^{\omega}&=\frac{M^2}{4\pi^2}\int_{\delta_{0}}^{\delta_{0}+\frac{2\pi}{M}}\int_{0}^{\frac{2\pi}{M}}E^{\theta_{A},\theta_{B},\frac{\omega}{2},\frac{\omega}{2}}_{0,0}Q^{\theta_{A},\theta_{B},\frac{\omega}{2},\frac{\omega}{2}}_{0,0}d\theta_{A}d\theta_{B}.\\
\end{aligned}
\end{equation}
Due to the random phase shifting, there is still an intrinsic QBER because the random phases are not perfectly matched. If $e_{\textrm{opt}}=0.03$, we have $\delta_{0}=0.35$ and $E_{XX}^{\nu}\sim3.6\%$.
By using the decoy-state mentod\cite{wang2005beating,lo2005decoy}, the yield $Y_{\textrm{TF}}^{1XX}$ and QBER $e_{XX}^{b1}$ can be given by
\begin{equation}
\begin{aligned}\label{}
Y_{\textrm{TF}}^{1XX}&\geq Y_{\textrm{TF}}^{1XXL}=\frac{\nu}{\nu  \omega -\omega ^2} \left(e^{\omega} Q^{\omega}_{XX}-\frac{ \omega ^2  }{\nu ^2}e^{\nu}Q^{\nu}_{XX}-\frac{\nu ^2-\omega ^2}{\nu ^2}Q^{0}\right),\\
e_{XX}^{b1}&\leq e_{XX}^{b1U}=\frac{e^\omega E_{XX}^{\omega}Q_{XX}^{\omega}-e^{b0}Q^{0}}{\omega Y_{\textrm{TF}}^{1XXL}},
\end{aligned}
\end{equation}
where $e^{b0}=\frac{1}{2}$ is the QBER of TF state with vacuum in phase-encoding basis.

For six-state encoding\cite{Lo2001Proof}, the probability that both bit flip and phase shift occurs can be given by\cite{yin2016security}
\begin{equation}
\begin{aligned}\label{}
a=(e_{ZZ}^{b1}+e_{XX}^{b1}-e_{YY}^{b1})/2.
\end{aligned}
\end{equation}
To simplify, we assume that those cases of qubit preparation with relative phase modulation are symmetrical since the random phase is unknown before Charlie performs single-photon BSM. Therefore, we abtain $a=e_{ZZ}^{b1}/2$. Interestingly, the QBER $e_{ZZ}^{b1}\equiv0$, which means that the key rate of practical TF-QKD with six-state encoding has no advantage compared with BB84 encoding.

For the RFI scheme\cite{laing2010reference}, the $Z$ basis is always well defined, which is $Z_{A}=Z_{B}=Z$ for Alice and Bob. The other two bases may vary with the slow phase shifting $\beta$, the relation can be given by $X_{B}=\cos \beta X_{A}+\sin \beta Y_{A}$,
$Y_{B}=\cos \beta Y_{A}-\sin \beta Y_{B}$ and $\beta=\beta_{B}-\beta_{A}$, where $Z_{A}$ and $Z_{B}$, $X_{A}$ and $X_{B}$, $Y_{A}$ and $Y_{B}$ are the location reference frames for $Z$, $X$ and $Y$ basis of Alice and Bob, respectively. $\beta_{A}$ ($\beta_{B}$) is the deviation between the practical and standard reference frame
for Alice (Bob). Therefore, the eigenstates of $X_{A}$ ($X_{B}$) and $Y_{A}$ ($Y_{B}$) can be written as $\ket{\pm}_{A}=(\ket{0}\pm e^{i\beta_{A}}\ket{1})/\sqrt{2}$ ($\ket{\pm}_{B}=(\ket{0}\pm e^{i\beta_{B}}\ket{1})/\sqrt{2}$) and $\ket{\pm i}_{A}=(\ket{0}\pm i e^{i\beta_{A}}\ket{1})/\sqrt{2}$ ($\ket{\pm i}_{B}=(\ket{0}\pm i e^{i\beta_{B}}\ket{1})/\sqrt{2}$). Note that $\beta_{A}$ and $\beta_{B}$ are the phases of intrinsic degree of freedom between $\ket{0}$ and $\ket{1}$ and can vary slowly in the virtual protocol with RFI theory.
The key rate of single-photon with RFI theory is given by\cite{laing2010reference}
\begin{equation}
\begin{aligned}\label{}
R_{\textrm{RFI}}=1-H(e_{b})-I_{E}(C).
\end{aligned}
\end{equation}
Here, $I_{E}(C)=(1-e_{b})H(\frac{1+\mu}{2})+e_{b}H(\frac{1+v}{2})$ quantifies the information of Eve's knowledge, parameters $v=\sqrt{C/2-(1-e_{b})^{2}u^2}/e_{b}$ and $u=\textrm{min}[\sqrt{C/2}/(1-e_{b}),1]$. We have $I_{E}(C)=H((1+\sqrt{C/2})/2)$ if the QBER $e_{b}=0$. The value $C$ can be defined as
\begin{equation}
\begin{aligned}\label{}
C=&\langle X_{A}X_{B}\rangle^2+\langle X_{A}Y_{B}\rangle^2+\langle Y_{A}X_{B}\rangle^2+\langle Y_{A}Y_{B}\rangle^2\\
=&(1-2E_{XX})^2+(1-2E_{XY})^2+(1-2E_{YX})^2+(1-2E_{YY})^2,
\end{aligned}
\end{equation}
which is independent of phase drifting $\beta_{A}$ ($\beta_{B}$) and can just be used to bound Eve's information. However, the phase drifting will add the QBER of $X$ basis, which will decrease the key rate of BB84 encoding. Thereinto, $E_{XX(YY,XY,YX)}$ is the QBER given that Alice and Bob choose $X-X(Y-Y,X-Y,Y-X)$ basis, which can be written as
\begin{equation}
\begin{aligned}\label{}
E_{XX}&=E_{YY}=\frac{1}{2}(1-\cos\beta),\\
E_{XY}&=\frac{1}{2}(1+\sin\beta),~~E_{YX}=\frac{1}{2}(1-\sin\beta).
\end{aligned}
\end{equation}
One can acquire the maximum value $C=2$ in the ideal case and $I_{E}(C=2)=0$ if the phase difference $\beta$ is fixed. For phase change from $\beta$ to $\beta+\Delta\beta$, $\Delta\beta\in[0,2\pi]$ (uniformity variation), we have
\begin{equation}
\begin{aligned}\label{}
C=\frac{2}{(\Delta\beta)^2}\left\{[\sin(\beta+\Delta\beta)-\sin\beta]^2+[\cos(\beta+\Delta\beta)-\cos\beta]^2\right\}=\frac{4(1-\cos\Delta\beta)}{(\Delta\beta)^2}.
\end{aligned}
\end{equation}
We can see that $C$ is only related to phase change $\Delta\beta$ and is not related to phase difference $\beta$ in theory. The value $C$ will decrease with $\Delta\beta$ increasing.

In the practical TF-QKD with RFI scheme, we define that $Q_{XXk}^{\nu}$ and $E_{XXk}^{\nu}$ are gain and QBER when Alice chooses $X$ basis with intensity $\frac{\nu}{2}$ and Bob chooses $X$ basis with intensity $\frac{\nu}{2}$ in the case of set $D_{k}$ by using the post-selected phase-matching method.
Therefore, the gain $Q_{XXk}^{\nu}$, $Q_{XYk}^{\nu}$, $Q_{YXk}^{\nu}$ and $Q_{YYk}^{\nu}$ of set $D_{k}$ are
\begin{equation}
\begin{aligned}\label{}
Q_{XXk}^{\nu}&=\frac{M^2}{4\pi^2}\int_{\delta_{0}+\frac{2\pi}{M}k}^{\delta_{0}+\frac{2\pi}{M}(k+1)}\int_{0}^{\frac{2\pi}{M}}Q^{\theta_{A},\theta_{B},\frac{\nu}{2},\frac{\nu}{2}}_{0,0}d\theta_{A}d\theta_{B},\\
Q_{XYk}^{\nu}&=\frac{M^2}{4\pi^2}\int_{\delta_{0}+\frac{2\pi}{M}k}^{\delta_{0}+\frac{2\pi}{M}(k+1)}\int_{0}^{\frac{2\pi}{M}}Q^{\theta_{A},\theta_{B},\frac{\nu}{2},\frac{\nu}{2}}_{0,1}d\theta_{A}d\theta_{B},\\
Q_{YXk}^{\nu}&=\frac{M^2}{4\pi^2}\int_{\delta_{0}+\frac{2\pi}{M}k}^{\delta_{0}+\frac{2\pi}{M}(k+1)}\int_{0}^{\frac{2\pi}{M}}Q^{\theta_{A},\theta_{B},\frac{\nu}{2},\frac{\nu}{2}}_{1,0}d\theta_{A}d\theta_{B},\\
Q_{YYk}^{\nu}&=\frac{M^2}{4\pi^2}\int_{\delta_{0}+\frac{2\pi}{M}k}^{\delta_{0}+\frac{2\pi}{M}(k+1)}\int_{0}^{\frac{2\pi}{M}}Q^{\theta_{A},\theta_{B},\frac{\nu}{2},\frac{\nu}{2}}_{1,1}d\theta_{A}d\theta_{B}.
\end{aligned}
\end{equation}
The QBER $E_{XXk}^{\nu}$, $E_{XYk}^{\nu}$, $E_{YXk}^{\nu}$ and $E_{YYk}^{\nu}$ of set $D_{k}$ can be written as
\begin{equation}
\begin{aligned}\label{}
E_{XXk}^{\nu}Q_{XXk}^{\nu}&=\frac{M^2}{4\pi^2}\int_{\delta_{0}+\frac{2\pi}{M}k}^{\delta_{0}+\frac{2\pi}{M}(k+1)}\int_{0}^{\frac{2\pi}{M}}E^{\theta_{A},\theta_{B},\frac{\nu}{2},\frac{\nu}{2}}_{0,0}Q^{\theta_{A},\theta_{B},\frac{\nu}{2},\frac{\nu}{2}}_{0,0}d\theta_{A}d\theta_{B},\\
E_{XYk}^{\nu}Q_{XYk}^{\nu}&=\frac{M^2}{4\pi^2}\int_{\delta_{0}+\frac{2\pi}{M}k}^{\delta_{0}+\frac{2\pi}{M}(k+1)}\int_{0}^{\frac{2\pi}{M}}E^{\theta_{A},\theta_{B},\frac{\nu}{2},\frac{\nu}{2}}_{0,1}Q^{\theta_{A},\theta_{B},\frac{\nu}{2},\frac{\nu}{2}}_{0,1}d\theta_{A}d\theta_{B},\\
E_{YXk}^{\nu}Q_{YXk}^{\nu}&=\frac{M^2}{4\pi^2}\int_{\delta_{0}+\frac{2\pi}{M}k}^{\delta_{0}+\frac{2\pi}{M}(k+1)}\int_{0}^{\frac{2\pi}{M}}E^{\theta_{A},\theta_{B},\frac{\nu}{2},\frac{\nu}{2}}_{1,0}Q^{\theta_{A},\theta_{B},\frac{\nu}{2},\frac{\nu}{2}}_{1,0}d\theta_{A}d\theta_{B},\\
E_{YYk}^{\nu}Q_{YYk}^{\nu}&=\frac{M^2}{4\pi^2}\int_{\delta_{0}+\frac{2\pi}{M}k}^{\delta_{0}+\frac{2\pi}{M}(k+1)}\int_{0}^{\frac{2\pi}{M}}E^{\theta_{A},\theta_{B},\frac{\nu}{2},\frac{\nu}{2}}_{1,1}Q^{\theta_{A},\theta_{B},\frac{\nu}{2},\frac{\nu}{2}}_{1,1}d\theta_{A}d\theta_{B}.
\end{aligned}
\end{equation}
By using the decoy-state method, the lower and upper bounds of yield $Y_{\textrm{TF}}^{1XXk}$, $Y_{\textrm{TF}}^{1XYk}$, $Y_{\textrm{TF}}^{1YXk}$ and $Y_{\textrm{TF}}^{1YYk}$ will be
\begin{equation}
\begin{aligned}\label{}
Y_{\textrm{TF}}^{1XXk}&\geq Y_{\textrm{TF}}^{1XXkL}=\frac{\nu}{\nu  \omega -\omega ^2} \left(e^{\omega} Q^{\omega}_{XXk}-\frac{ \omega ^2  }{\nu ^2}e^{\nu}Q^{\nu}_{XXk}-\frac{\nu ^2-\omega ^2}{\nu ^2}Q^{0}\right),\\
Y_{\textrm{TF}}^{1XYk}&\geq Y_{\textrm{TF}}^{1XYkL}=\frac{\nu}{\nu  \omega -\omega ^2} \left(e^{\omega} Q^{\omega}_{XYk}-\frac{ \omega ^2  }{\nu ^2}e^{\nu}Q^{\nu}_{XYk}-\frac{\nu ^2-\omega ^2}{\nu ^2}Q^{0}\right),\\
Y_{\textrm{TF}}^{1YXk}&\geq Y_{\textrm{TF}}^{1YXkL}=\frac{\nu}{\nu  \omega -\omega ^2} \left(e^{\omega} Q^{\omega}_{YXk}-\frac{ \omega ^2  }{\nu ^2}e^{\nu}Q^{\nu}_{YXk}-\frac{\nu ^2-\omega ^2}{\nu ^2}Q^{0}\right),\\
Y_{\textrm{TF}}^{1YYk}&\geq Y_{\textrm{TF}}^{1YYkL}=\frac{\nu}{\nu  \omega -\omega ^2} \left(e^{\omega} Q^{\omega}_{YYk}-\frac{ \omega ^2  }{\nu ^2}e^{\nu}Q^{\nu}_{YYk}-\frac{\nu ^2-\omega ^2}{\nu ^2}Q^{0}\right),\\
\end{aligned}
\end{equation}
and
\begin{equation}
\begin{aligned}\label{}
Y_{\textrm{TF}}^{1XXk}&\leq Y_{\textrm{TF}}^{1XXkU}=\frac{e^\omega Q_{XXk}^{\omega}-Q^{0}}{\omega},~~~~Y_{\textrm{TF}}^{1XYk}\leq Y_{\textrm{TF}}^{1XYkU}=\frac{e^\omega Q_{XYk}^{\omega}-Q^{0}}{\omega},\\
Y_{\textrm{TF}}^{1YXk}&\leq Y_{\textrm{TF}}^{1YXkU}=\frac{e^\omega Q_{YXk}^{\omega}-Q^{0}}{\omega},~~~~Y_{\textrm{TF}}^{1YYk}\leq Y_{\textrm{TF}}^{1YYkU}=\frac{e^\omega Q_{YYk}^{\omega}-Q^{0}}{\omega}.
\end{aligned}
\end{equation}
The lower and upper bounds of QBER $e_{XXk}^{b1}$, $e_{XYk}^{b1}$, $e_{YXk}^{b1}$ and $e_{YYk}^{b1}$ can be given by
\begin{equation}
\begin{aligned}\label{}
e_{XXk}^{b1}&\geq e_{XXk}^{b1L}=\frac{\nu}{(\nu  \omega -\omega ^2)Y_{\textrm{TF}}^{1XXkU}} \left(e^{\omega} E^{\omega}_{XXk}Q^{\omega}_{XXk}-\frac{ \omega ^2  }{\nu ^2}e^{\nu}E^{\nu}_{XXk}Q^{\nu}_{XXk}-\frac{\nu ^2-\omega ^2}{\nu ^2}e^{b0}Q^{0}\right),\\
e_{XYk}^{b1}&\geq e_{XYk}^{b1L}=\frac{\nu}{(\nu  \omega -\omega ^2)Y_{\textrm{TF}}^{1XYkU}} \left(e^{\omega} E^{\omega}_{XYk}Q^{\omega}_{XYk}-\frac{ \omega ^2  }{\nu ^2}e^{\nu}E^{\nu}_{XYk}Q^{\nu}_{XYk}-\frac{\nu ^2-\omega ^2}{\nu ^2}e^{b0}Q^{0}\right),\\
e_{YXk}^{b1}&\geq e_{YXk}^{b1L}=\frac{\nu}{(\nu  \omega -\omega ^2)Y_{\textrm{TF}}^{1YXkU}} \left(e^{\omega} E^{\omega}_{YXk}Q^{\omega}_{YXk}-\frac{ \omega ^2  }{\nu ^2}e^{\nu}E^{\nu}_{YXk}Q^{\nu}_{YXk}-\frac{\nu ^2-\omega ^2}{\nu ^2}e^{b0}Q^{0}\right),\\
e_{YYk}^{b1}&\geq e_{YYk}^{b1L}=\frac{\nu}{(\nu  \omega -\omega ^2)Y_{\textrm{TF}}^{1YYkU}} \left(e^{\omega} E^{\omega}_{YYk}Q^{\omega}_{YYk}-\frac{ \omega ^2  }{\nu ^2}e^{\nu}E^{\nu}_{YYk}Q^{\nu}_{YYk}-\frac{\nu ^2-\omega ^2}{\nu ^2}e^{b0}Q^{0}\right),\\
\end{aligned}
\end{equation}
and
\begin{equation}
\begin{aligned}\label{}
e_{XXk}^{b1}&\leq e_{XXk}^{b1U}=\frac{e^\omega E_{XXk}^{\omega}Q_{XXk}^{\omega}-e^{b0}Q^{0}}{\omega Y_{\textrm{TF}}^{1XXkL}},~~~~~e_{XYk}^{b1}\leq e_{XYk}^{b1U}=\frac{e^\omega E_{XYk}^{\omega}Q_{XYk}^{\omega}-e^{b0}Q^{0}}{\omega Y_{\textrm{TF}}^{1XYkL}},\\
e_{YXk}^{b1}&\leq e_{YXk}^{b1U}=\frac{e^\omega E_{YXk}^{\omega}Q_{YXk}^{\omega}-e^{b0}Q^{0}}{\omega Y_{\textrm{TF}}^{1YXkL}},~~~~~e_{YYk}^{b1}\leq e_{YYk}^{b1U}=\frac{e^\omega E_{YYk}^{\omega}Q_{YYk}^{\omega}-e^{b0}Q^{0}}{\omega Y_{\textrm{TF}}^{1YYkL}}.\\
\end{aligned}
\end{equation}
For the practical TF-QKD with RFI scheme, we need to calculate the minimum value of $C_{k}^{1}$. Therefore, for the value
\begin{equation}
\begin{aligned}\label{}
C_{k}^{1}=(1-2e_{XXk}^{b1})^2+(1-2e_{XYk}^{b1})^2+(1-2e_{YXk}^{b1})^2+(1-2e_{YYk}^{b1})^2,
\end{aligned}
\end{equation}
we have
\begin{equation}
e_{XXk}^{b1}=\left\{
\begin{array}{rcl}
e_{XXk}^{b1U},& & {e_{XXk}^{b1U}\leq\frac{1}{2}},\\
e_{XXk}^{b1L}, & & {e_{XXk}^{b1L}\geq\frac{1}{2}},\\
\frac{1}{2}, & & {e_{XXk}^{b1L}\leq\frac{1}{2}\leq e_{XXk}^{b1U}},\\
\end{array} \right.
\end{equation}
the parameters $e_{XYk}^{b1}$, $e_{YXk}^{b1}$ and $e_{YYk}^{b1}$ are similar with the case of $e_{XXk}^{b1}$.

\noindent\textbf{Acknowledgments}\\
We thank W. Zhu for the help with the figure.
H.-L. Yin gratefully acknowledges support from the National Natural Science Foundation of China under Grant No. 61801420, the Fundamental Research Funds for the Central Universities.

\noindent\textbf{Author Contributions}\\
H.-L.Y. and Y.F. have the main idea. All results are acquired through the
discussion among all authors. All authors contribute to the writing and reviewing of the manuscript.

\noindent\textbf{Additional Information}\\
Competing interests: The authors declare no competing interests.


\bibliographystyle{naturemag}


\clearpage

\clearpage

\clearpage

\clearpage

\end{document}